# Towards Simplifying PKI Implementation: Client-Server based Validation of Public Key Certificates

Diana Berbecaru    Antonio Lioy

Dip. Automatica e Informatica

Politecnico di Torino,

Corso Duca degli Abruzzi, No. 24, 10129, Torino, Italy

**Abstract**

With real-time certificate validation checking, a public-key-using system that needs to validate a certificate executes a transaction with a specialized validation party. At the end of the transaction the validation party returns an indication about the validity status of the certificate. This paper analysis the public key (PbK) certificate validation service from a practical point of view by describing the implementation of a system that makes use of the Data Validation and Certification Server (DVCS) protocols to provide certificate validation service to the Relying Parties (RPs). However the system is not restricted to use only the specified protocol and allows the integration of other validation protocols or mechanisms. Our implementation efforts emphasize the possibility to pursue a specific RP tradeoff between timeliness, security and computational resource usage via dynamic selection of several configurable options.
**Keywords**: PKI, certificate validation, DVCS, validation options.

## 1 INTRODUCTION

Despite widely recognized importance of Public Key Infrastructure (PKI), significant design issues that impede the scalability and interoperability among large scale PKIs are still open. One particularly debated aspect is the mechanism for validating PbK certificates. None of the mechanisms proposed so far for certificate validation (CV) could alone meet the computational, network and security requirements of all relying party applications and PKI topologies. CV is a complex process and requires the construction and validation of certification paths. Shortly, this is named also certificate path processing. A certification path is said to be valid under a set of certificate policies that are common to all certificates in the path [1]. Certificate path processing must be implemented in every certificate-using system or in a server that supports that system.

The IETF PKIX working group defined the requirements [2] and is currently examining validation protocols [3] [4] [5] used in client-server architectures where the client can delegate the path processing to a server with the scope to minimize the relying party's client complexity. In these schemes certificate path processing is passed partially or totally from the RP to a dedicated server, further called Certificate Validation Server (CVS). The protocols deal with certification paths processing against a number of CVS constraints called *validation policies*. To perform its task, the CVS extracts data from a number of resources (e.g. directories, local repositories) and communicates with various external servers.

This paper describes first a **S**ecure **A**rchitecture for the **Va**lidation of X.509v3 **Cert**ificates (SAVaCert) based on the client-server model and designed at a suitable level of abstraction. General functionality is defined for the component modules to allow implementers to choose particular protocols and validation mechanisms on the client and server sides. Once a specific protocol is chosen, it is useful to give to relying parties the possibility to dynamically select (depending on their current context) the most appropriate validation parameters and send them to the server. In other words, SAVaCert allows clients to specify different validation profiles in accordance with their timeliness, security and power/network resources requirements. Additionally, the user has the possibility to send to the server either the exact input certificate policy data to be used by the CVS during path validation or it can give vague indications, like its intended usage of the certificate, by using the method explained in Section 2. Section 3 describes our implementation of SAVaCert based on the use of DVCS protocols [5] and of existent implementations for the modules on the server.

SAVaCert is addressed to the following categories of users:
**a. thin** and **off-line users**, i.e. users that are scarce in computational and respectively network resources.
**b. enterprise environments,** i.e. environments where the policy control and management of trust is performed centrally. The security administrators can configure and maintain policy data and trusted anchors on a dedicated CVS and the enterprise users can delegate the certificate path processing to a trusted CVS. Most of the users have little or no idea about security policies and operating practices of various CAs, about retrieving the freshest certificate status (CS) information for each certificate in the path or about securely obtaining, storing and maintaining the *most-trusted CAs* [6] [7]. These clients trust the CVS to extract the necessary data and perform the path processing.

*c. simplified users* need assistance from the CVS to minimize the risk of using invalid certificates. This category of users would require CVS to evaluate the certificate policies and construct the certificate chains or to retrieve on-line responses (called OCSP responses [8]) containing the revocation status of a certificate. In this case the CVS is not necessarily trusted.

## 2 DESCRIPTION OF SAVaCert

The purpose of this section is the description of the architecture for PbK certificate validation at a suitable level of abstraction. The main components of the architecture and their functionality are described such that to permit a number of implementation possibilities. In SAVaCert the client delegates subtasks (e.g. only path discovery) or the entire task (e.g. path discovery and path validation) of certificate path processing to a server, as it is depicted in Figure 1.

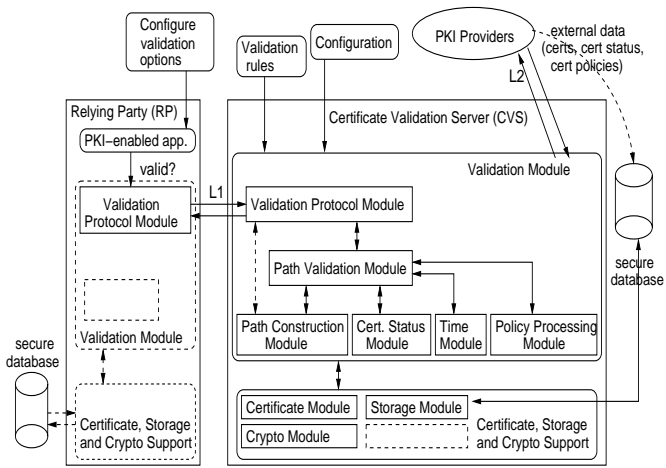

Figure 1: SAVaCert Components

We'll recall shortly the main steps to be performed in order to validate a certificate, called also *targetCert*:
*a. path construction.* This may require a path discovery task resulting in several certification paths found by the CVS for a certain *targetCert* .
*b.* execution of a ***path validation algorithm*** that includes certificate verification (i.e. whether it has expired or is revoked) for each certificate in the path and the processing of path constraints. The algorithm must verify also the digital signature on each certificate, check that the required certificate policies are indicated in the certificates and check that the names in the certificates are consistent with a valid certification path, that is, the subject of every certificate in the path is the issuer of the next certificate (except the root CA). Besides the *targetCert*, the request sent on L1 can contain optionally data such as certificate revocation data or policy information. The RP relies on the CVS to perform a service constrained by the so-called *validation policy*, under which the validation is requested and that the RP considers acceptable for one PKI-enabled application. CVS constructs a response containing either the results of verifications for the *targetCert* or an error message. Ford and Baum [9] note that it is helpful to consider the certification path discovery and validation as two separate tasks. Their implementations may be separate because the certification path discovery is not a security-critical function whereas the path validation is.

In this model the *Validation Module* of the CVS is not viewed as an ad-hoc application but has been further split in several components each of which has a dedicated task as described next:
*Validation Protocol Module (VPM)*. The main role of this module is to manage the messages exchanged on link L1. The module contains functions to create, print, sign, parse requests and responses.
*Path Validation Module (PVM)*. Its main role is to execute path validation to determine the validation status one *targetCert*. The algorithms for path validation are included in this abstract module.
*Path Construction Module (PCM)*. The role of this module is to construct certification paths. Efficient algorithms for path development, which use certificate *extensions* and loop detection/elimination techniques to increase efficiency [10] [11], are embedded at this abstract level.
*Certificate Status Module (CSM)* has to determine the status of a certificate. The code for retrieving and processing CRLs is contained in this module. If required, an OCSP client is integrated here to communicate with various OCSP responders. Support for other revocation mechanisms, like for example Certificate Revocation Trees (CRTs) or authenticated dictionaries, is included also in this module.
*Policy Processing Module (PPM)*. The role of this module is to process policies. This module is divided further in submodules, one of them processes the certificate policies and the other one processes the validation policies.
*Time module (TM)*. The role of this module is to give an indication of time. The simplest form is time represented as GeneralizedTime but alternate forms include a TST [12] or a DVC [5].

The *Storage Module* is used to store/retrieve certificates, certificate revocation data and policies. Each module accesses the other modules through interfaces. A typical server application receives and parses the request by making calls to *VPM*. The values contained in the request fields are passed as input to the *PVM* or *PCM*, which will make further calls to functions from the other modules. The *Validation Module* is not based on a particular cryptographic library or certificate management library to allow developers to freely choose the underlying support. Since certificates and all PKIX protocols are specified using Abstract Syntax Notation 1 (ASN.1), a component will be also required to perform ASN.1 parsing and generation. This component is also considered in SAVaCert as being part of the underlying support.

This is a general architecture, void of details on the use of a specific protocol. Thus, L1 and L2 can be mapped onto particular protocols of choice. CVS uses L2 to update the secure database and to communicate with other TTPs. Not all modules are compulsory. If the task of the CVS is to perform path discovery then only the *VPM*, *PCM* and sub modules from the *PPM* are present while the others are optional. Similarly the RP can contain a variable number of modules. For example *thin* clients [13] would require the *VPM* and optionally the support for digital signing.

In SAVaCert the RP can configure several parameters to control or restrict the validation process. We have divided them into the following distinct categories:

*Validation Parameters.* The RP must have a mean to indicate to the CVS how much information used during path processing must be included into the response, e.g. PbK certificates, CRLs, OCSP responses, validation time. The RP must be able also to verify finally that the CVS has made the verifications using the validation parameters accordingly.

*Certificate Policy Processing Parameters.* When validating one *targetCert*, the *PPM* must examine the certificate policies of the CAs from one certification path. The RP sets these parameters to specify its strict or weak certificate policy requirement (CPR) to the CVS. The *strict CPR* is used when the RP knows the set of certificate policies that are acceptable for its intended use. Additionally, the RP can give an indication to the CVS about how to process the policy constraints. In this case the CVS can give an *on/off* response, i.e. it will indicate the path to be valid only if it is valid under one of the certificate policies specified in the set and it would return the subset for which the path is valid. Instead the RPs specify *weak CPR* to give a general indication about the intended usage of the certificate (e.g. e-mail) when they don't know which policies are accept-able for their use and consequently they ask the CVS to build and validate the appropriate certification path(s).

*Validation Protocol Module (VPM) Parameters* are characteristic to the protocol chosen for the link L1.

## 3 A PRACTICAL APPROACH

The IETF PKIX working group defined the Data Validation and Certification Server (DVCS) and the protocols to be used when communicating with it. The role of the server is to assert the validity of signed documents, public key certificates and the possession or existence of data. From these functionalities we were particularly interested in the service that allows a client to offload certificate validation to a server. In this case, SAVaCert consists basically of a DVCS server that makes use of various data sources and communicates with various servers to provide signed response structures called Data Validation Certificates (DVCs) to the DVCS clients. The DVCs contains validation results and trustworthy time information that asserts the validity status of PbK certificates.

A DVCS transaction begins with a client preparing a request that contains data for which validity is to be certified, data to be used by the server during the verification processes and extensions to carry additional data. The clients can ask the server to validate more than one certificate at once. Thus, all the *targetCerts*, the chains (if any) and the policy information are embedded into one single request. The client prepares the request by making calls to the *VPM*. The client can control the validation process by setting several parameters with a simple interface as illustrated in Figure 2.

Figure 2: Sample DVCS client profile

At this step the client choose to use *strict* or *weak CPR*. In case of *strict CPR* the client can specify an *acceptable policy set* consisting of one or more acceptable certificate policies identifiers which the VPM inserts in the request in the *acceptable policy set*. For example, the client may choose to accept only paths consistent with some high-assurance certificate policy identifier. The *PPM* on the server will indicate the path to be valid only if it is valid under one or more of the policies specified in this set and it would return the subset of the *acceptable policy set* for which the path is valid.

Additionally, the client can give indications to the server about how to process the certificate policy OID extensions. Thus, two boolean flags in the request, an *explicit policy required* and *inhibit policy mapping* (whose meaning is identical to the fields in the *Policy Constraints* [6]) can be set to indicate further to the server whether or not the policy mapping is allowed and to explicitly specify the policies required. In this context their effect is to allow the client to set the policy constraints state immediately, without waiting for the certificate extension to appear in the path. The *explicit policy required* determines if an acceptable policy identifier as defined in [6] needs to explicitly appear in the certificate policies *extension* field of all certificates in the path. By setting to *true* this flag, the client indicates that the certificate policies in the chain must be consistent with the *acceptable policy set*. The *inhibit policy mapping* determines if policy mapping will be inhibited or not during path processing. By setting this flag to *true* the client indicates to the server that the identifiers in the *acceptable policy set* must not be mapped by a CA in the chain. If the client chooses the *strict CPR* but it doesn't specify an *acceptable policy set* and the flags then the fields of the request used for storing these values are left blank. This will be interpreted by the CVS that *any policy* is acceptable for the client. In the other case, if the client chooses the *weak CPR*, the VMP doesn't insert specific certificate policy data but rather the client's general criterion is inserted as *extension* in the request. We note that the absence of certificate policy data in the request (*acceptable policy set* and flags) must not lead the server to the conclusion that *any policy* is acceptable if the *extension* of the request contains some data. When the server constructs the response it will set the *acceptable policy set* to indicate the set of certificates for which the certification path is valid and the policy mappings that were processed. If the *Default* option is chosen in the Figure 2 as weak CPR then this means that *any policy* is acceptable for the client.

*Validation parameters* help a client to indicate the validation policy under which the validation is requested and under which the DVCS server operates. The client can express this policy with an identifier that the VPM inserts into the request in the *requestPolicy* field of the *requestInformation* structure. This case can be applied if the client knows in advance both the validation policy that is appropriate for one particular application as well as its unique identifier. Another possibility would be either to map the validation options to an identifier or to define *extensions* in the request (possibly also in the *requestInformation* structure) to hold the validation options but in this case the server must be able to process these *extensions*. The client can verify that the server has made the verification in accordance to the validation parameters because a copy of the *requestInformation* of the corresponding request is inserted in the response.

The client can set also a number of VPM Parameters due to the use of DVCS for L1, such as: check the validity of the server's signing certificate using on-line or off-line mechanisms; store part of the information for later use; trust or not the unsigned responses (this may be the case if the client uses a transport mechanism that provides server authentication); require CMS encapsulation [14] for the request or to establish an SSL channel [15] to ensure not only server authentication but also the confidentiality of data exchanged.

When it receives one request the server checks first if the request is valid, e.g. whether it contains an acceptable time, the correct name of the server, the correct request information and the appropriate service. The *VPM* extracts from the request one *targetCert* and passes it to the *PVM*. The certificates and algorithms for validation of certification paths discussed in this paper will respect the conventions of X.509 [1] and the PKIX [6] [7] profile even though, according to the above mentioned standard, any algorithm could be implemented as long as the results are guaranteed to be the same as these standard algorithms.

If more than one *targetCert* is present in the request then they will be processed sequentially. Unless the client has inserted also a set of certificates in the request to be used for the construction of the certification paths, or if the server's local policy does not allow to use this set, the *PVM* calls functions from the *PCM* in order to execute a path discovery process. Other *extensions* of the request can be defined to support the certificate path construction on the server but they are beyond the scope of this paper. The path discovery will result in a number of possible 'valid' chains called *candidate chains*. To choose the correct chain the *PCM* needs to receive from the *VPM* the values of the request fields that contain either the policy OIDs and policy constraints flags or the *extension* that contain a client's vague criterion. In other words, the server can determine the certificate policy processing method chosen by the client only after it retrieves the fields in the request that carry the certificate policy data and the appropriate re-quest *extension*. If none of the above data is present in the request then the server concludes that the client considers acceptable *any policy*. The server cannot determine exactly at this point which paths are good without doing a path validation also. Thus, one candidate chain is passed as input to the *PVM* to be processed. Throughout the path validation process the name constraints are checked also and function calls are made to the *PPM* to check the policy constraints. If required, functions from *CSM* and *TM* are called to return OCSP or TSP [12] responses. After completion of path validation there could still be more than one path satisfying the client's criteria. In the simplest case, which is also the most common, the client is asking for a single chain. Finally, the server *PVM* returns the validation results and the other related data (like the set of policies used and the mappings that were processed during the execution of certificate path validation algorithm) to the *VPM* which will construct a DVC or an error message. When the client receives the DVC, it will verify first that the server signing certificate is valid if the client has configured this option. The client will interpret further the response. For example, if validation for one certificate failed because of revocation, the client presents the CRL in a GUI display.

The first step in implementing SAVaCert was the implementation of the *VPM* to generate, parse, sign and verify DVCS messages by making use of OpenSSL [16]. For the DER [17] functions the Valicert parser [18] was used and the code has been slightly modified afterwards. To enable OCSP functionality we've embedded an OCSP client into the *CSM* and the calls to this module are made via a simple OCSP client API [19]. For the PCM and PVM we've used the Certificate Path Development Library [20] and respectively the Certificate Management Library [21]. Other freeware implementations for building/validating certificate chains are available and may be used, such as IBM's Jonah PKIX reference implementation [22] or Pequi [23]. To support the weak CPR we chose a secure database whose physical format allows the storage of an indication of the purposes for which one certificate is *trusted* [24].

## 4 CONCLUSIONS AND FUTURE WORK

Certificate validation is an important process because if it were done in an incorrect manner or incompletely an end entity would use the certificate deemed to be valid. This paper describes the components and the functionality of a general modular architecture for certificate validation named SAVaCert. Further our implementation of SAVaCert was described. The system is based on the client-server model and the DVCS protocol. Nevertheless clients and servers are not restricted in SAVaCert to use particular protocols. Moreover clients can specify parameters to control the validation process. Our future work will be focused on evaluating the complexity of the server and comparison with other proposed certificate validation protocols.